\documentclass[letterpaper,twocolumn,english,prl,superscriptaddress]{revtex4-1}
\usepackage{amsmath}
\usepackage{newtxmath}
\usepackage[T1]{fontenc}
\usepackage[latin9]{inputenc}
\usepackage{color}
\usepackage{babel}
\usepackage{graphicx}
\usepackage[pdfusetitle,
 bookmarks=true,bookmarksnumbered=false,bookmarksopen=false,
 breaklinks=false,pdfborder={0 0 1},backref=false,colorlinks=true]
 {hyperref}

\makeatletter

\usepackage{babel}

\makeatother

\begin{document}
\title{Stable excitations and holographic transportation in tensor networks
of critical spin chains}
\author{Zuo Wang}
\affiliation{Institute for Theoretical Physics, School of Physics, South China
Normal University, Guangzhou 510006, China}
\affiliation{Key Laboratory of Atomic and Subatomic Structure and Quantum Control
(Ministry of Education), Guangdong Basic Research Center of Excellence
for Structure and Fundamental Interactions of Matter, School of Physics,
South China Normal University, Guangzhou 510006, China}
\affiliation{Guangdong Provincial Key Laboratory of Quantum Engineering and Quantum
Materials, Guangdong-Hong Kong Joint Laboratory of Quantum Matter,
South China Normal University, Guangzhou 510006, China}
\author{Liang He}
\email{liang.he@scnu.edu.cn}

\affiliation{Institute for Theoretical Physics, School of Physics, South China
Normal University, Guangzhou 510006, China}
\affiliation{Key Laboratory of Atomic and Subatomic Structure and Quantum Control
(Ministry of Education), Guangdong Basic Research Center of Excellence
for Structure and Fundamental Interactions of Matter, School of Physics,
South China Normal University, Guangzhou 510006, China}
\affiliation{Guangdong Provincial Key Laboratory of Quantum Engineering and Quantum
Materials, Guangdong-Hong Kong Joint Laboratory of Quantum Matter,
South China Normal University, Guangzhou 510006, China}
\begin{abstract}
The AdS/CFT correspondence conjectures a duality between quantum gravity
theories in anti-de Sitter (AdS) spacetime and conformal field theories
(CFTs) on the boundary. One intriguing aspect of this correspondence
is that it offers a pathway to explore quantum gravity through tabletop
experiments. Recently, a multi-scale entanglement renormalization
ansatz (MERA) model of AdS/CFT that can be implemented using contemporary
quantum simulators has been proposed {[}R. Sahay, M. D. Lukin, and
J. Cotler, arXiv:2401.13595 (2024){]}. Particularly, local bulk excitations
(entitled ``hologrons'') manifesting attractive interactions given
by AdS gravity were found. However, the fundamental question concerning
the stability of these identified hologrons is still left open. Here,
we address this question and find that hologrons are unstable during
dynamic evolution. In searching for stable bulk excitations with attractive
interactions, we find they can be constructed by the local primary
operators in the boundary CFT. Furthermore, we identify a class of
boundary excitations that exhibit the bizarre behavior of ``holographic
transportation'', which can be directly observed on the boundary
system implemented in experiments. 
\end{abstract}
\maketitle
\emph{Introduction}.---The AdS/CFT correspondence is a cornerstone
in the exploration of quantum gravity \cite{Maldacena1999IJTP,Gubser_PLB_1998,witten1998ATMP}.
It conjectures that the quantum gravity theory in a $(d+1)$-dimensional
anti-de Sitter (AdS) spacetime is equivalent to a conformal field
theory (CFT) without gravity on the $d$-dimensional boundary. This
duality provides a concrete realization of the holographic principle
in quantum gravity \cite{hooft_2009_arxiv,Susskind_1995_JMP}, with
a prominent example being the duality between type IIB superstring
theory on $\text{AdS}_{5}\times\text{S}^{5}$ and $\mathcal{N}=4$
super Yang-Mills theory in four dimensions \cite{Maldacena1999IJTP}.
The correspondence has also been extended to other dimensions, with
the $\text{AdS}_{3}/\text{CFT}_{2}$ case being particularly notable
due to the tractability of 2D CFTs, such as the 2D Ising CFT \cite{francesco2012Springer},
which facilitates non-perturbative investigations of quantum gravity
and black hole physics in $\text{AdS}_{3}$ through quantum simulations
of the boundary theory. These boundary theories, as low-energy effective
descriptions of critical quantum systems, are increasingly accessible
to experimental realization, thanks to recent advances in quantum
simulation platforms---such as trapped ions \cite{Monroe2021RMP, haghshenas_PRL_2024},
Rydberg atom arrays \cite{Keesling2019nature, Slagle2021PRB, fang2024arxiv},
and ultracold atoms in optical lattices \cite{kim2024arXiv}---which
are opening new avenues for exploring gravitational physics in a controlled
laboratory setting.

To bridge the gap between theoretical advancements and experimental
realizations, various models have been proposed, such as the SYK model
\cite{Sachdev_Ye_1993_PRL,Kitaev_2015_talk,Maldacena_Stanford_2016_PRD,Susskind_Swingle_2020_arxiv,Jafferis_2022_nature}
and the BFSS matrix model \cite{Susskind_1997_PRD,Maldacena_1998_PRD,Masanori_2014_science,Berkowitz_2016_PRD,Rinaldi_2022_PRX,Pateloudis_2023_JHEP,maldacena_2023_arxiv}.
While these models have attracted significant interest, their large-scale
simulations remain technically challenging \cite{Norman_Stanford_2021_PRL,garcia_2023_arxiv,maldacena_2023_arxiv}.
In contrast, tensor network approaches, particularly the multi-scale
entanglement renormalization ansatz (MERA), offer a promising framework
for exploring holographic mappings \cite{Swingle2012PRD,Vidal2008PRL,Vidal2007PRL}.
MERA provides an exact unitary mapping between a boundary CFT and
a bulk system \cite{Chua2017PRB,xiaoliang_2013_arxiv}, capturing
the AdS/CFT correspondence at the level of Hilbert spaces \cite{harlow_2018_lactures_arxiv}
and potentially yielding holographic dictionaries such as the GKPW
relation \cite{Gubser_PLB_1998,witten1998ATMP} and the BDHM relation
\cite{Banks_arXiv_1998}. Recently, Ref.~\cite{sahay2024arXiv} introduced
a MERA-based model mapping a ($1+1$)-dimensional critical spin chain
to a ($2+1$)-dimensional bulk spin system, unveiling \textquotedblleft hologrons\textquotedblright ,
massive bulk excitations that experience an attractive force akin
to gravity in AdS spacetime. This work establishes a compelling gravitational
analogy, sparking interest in whether such models can capture gravitational
dynamics, such as particle interactions and black hole dynamics. However,
the stability of these hologrons remains an open question, representing
a crucial step toward validating the model\textquoteright s capacity
to probe gravitational dynamics. If hologrons prove unstable, identifying
a stable quasiparticle becomes crucial. Moreover, it remains uncertain
whether the model captures a broader range of gravitational phenomena
and AdS/CFT physics beyond the observed attractive forces. 

In this work, we address the open questions related to the experimentally
accessible holographic model for AdS/CFT proposed in Ref.~\cite{sahay2024arXiv},
where a critical spin chain is unitarily mapped to a bulk MERA tensor
network, and obtain the following key results. (i) We show that hologrons
are dynamically unstable. (ii) We identify a class of dynamically
stable bulk quasiparticles, constructed from the local primary operators
of the boundary CFT. These quasiparticles are localized, massive,
and exhibit attractive interactions, therefore they could be good
candidates for exploring gravitational physics in holographic models.
(iii) We observe a bizarre phenomenon, termed \textquotedblleft holographic
transportation\textquotedblright , rooted in the AdS/CFT correspondence,
in which boundary excitations temporarily vanish from the boundary
and undergo a process analogous to being \textquotedblleft transported\textquotedblright{}
along AdS geodesics via the extra dimension of the holographic AdS
bulk. Moreover, we find that this holographic transportation, a phenomenon
deeply rooted in the AdS/CFT correspondence, is robust against typical
experimental imperfections, making it a highly promising candidate
for direct observation in state-of-the-art quantum simulation platforms,
such as programmable Rydberg atom arrays \cite{Bernien_Nature_2017,Keesling2019nature,Ebadi2021nature,Scholl2021nature,Semeghini_Science_2021,Bluvstein_2022_nature,sahay2024arXiv}.

\begin{figure}
\includegraphics[width=1.7in]{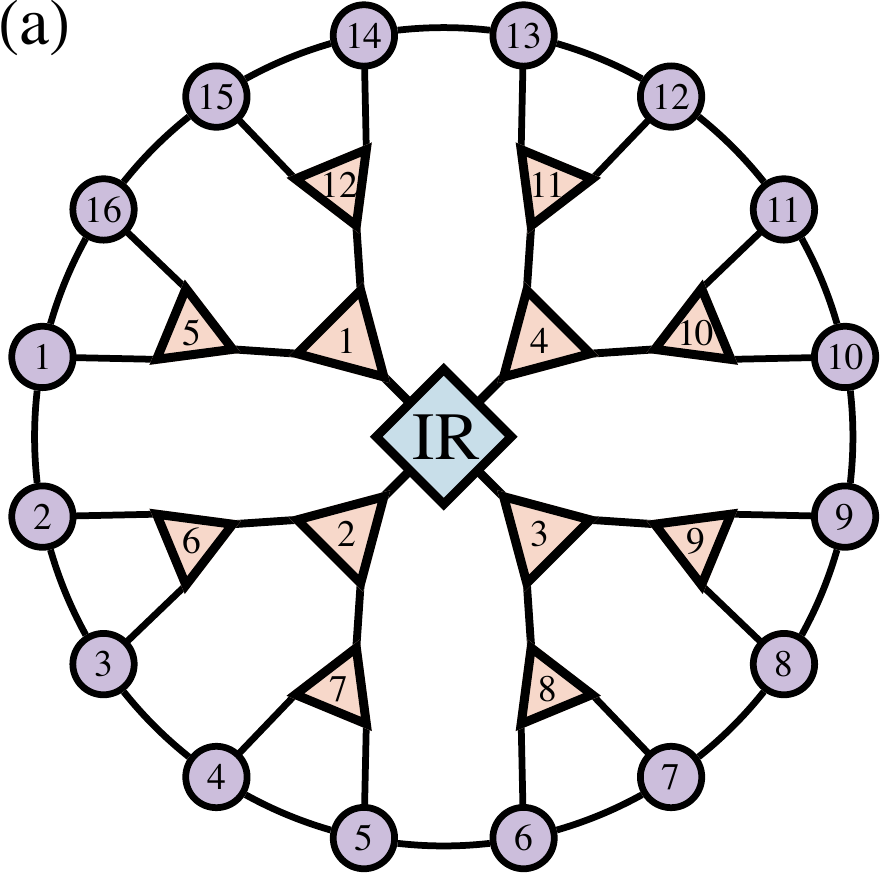}~~\includegraphics[width=1.5in]{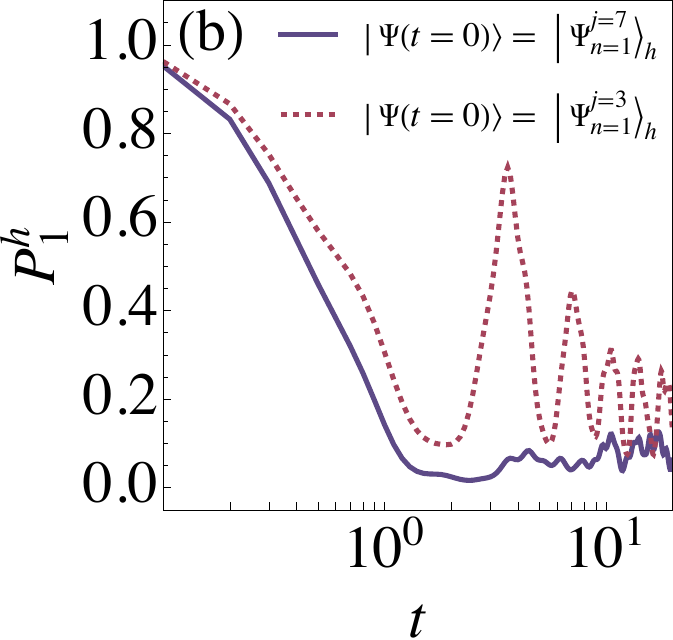}\caption{\label{fig: hologron_stability}Hologron stability in the MERA model.
(a) Physical degrees of freedom in the MERA tensor network \cite{sahay2024arXiv}
implemented in the simulation. The circles represent the spins on
the boundary critical spin chain, the triangles correspond to bulk
spins, and the square at the center indicates the infrared degree
of freedom of the bulk system. (b) Time evolution of $P_{1}^{h}(t)$
with initial bulk states $|\Psi_{n=1}^{j=7}\rangle_{h}$ and $|\Psi_{n=1}^{j=4}\rangle_{h}$.
See text for more details. }
\end{figure}

\emph{Hologrons and their instability}.---We consider the holographic
model proposed in Ref.~\cite{sahay2024arXiv} that is accessible to
experimental simulations on current quantum simulation platforms.
It consists of a unitary mapping between a ($1+1$)-dimensional critical
spin chain on the boundary and a ($2+1$)-dimensional ``bulk'' system
{[}see Fig.~\ref{fig: hologron_stability}(a) for a schematic illustration{]}.
More specifically, the critical spin chain on the boundary is described
by the Hamiltonian $H_{\partial}$, which takes the explicit form:
\begin{equation}
H_{\partial}=J\frac{L}{4\pi}\sum_{j=1}^{L}\left(X_{j-1}Z_{j}X_{j+1}-X_{j}X_{j+1}\right),\label{eq: bdyHam}
\end{equation}
where $X,Y,Z$ are Pauli operators, $J$ is the interaction energy,
$L$ is the total number of lattice sites, and periodic boundary conditions
are assumed. Throughout this study, we work in units where $J=1$
and $\hbar=1$. The bulk system consists of $L-4$ bulk spins {[}triangle
in Fig.~\ref{fig: hologron_stability}(a){]} and a core region {[}square
in Fig.~\ref{fig: hologron_stability}(a){]}. Its Hamiltonian $H$
is related to the boundary system $H_{\partial}$ via the holographic
unitary map $\mathcal{U}$, i.e., $H=\mathcal{U}^{\dagger}H_{\partial}\mathcal{U}$,
where $\mathcal{U}$ is derived from the MERA \cite{Vidal2007PRL,Vidal2008PRL,Evenbly2009PRB}
tensor network representation of the ground state of the critical
spin chain $H_{\partial}$. We use the class of scale-invariant MERA
\cite{Evenbly2016PRL}, which allows for an analytical expression
of $\mathcal{U}$ (see Supplemental Material (SM) \cite{Sup_Mat}
for technical details).

The ground state of the bulk system $|\Psi_{\text{GS}}\rangle$ takes
the form $|\Psi_{\text{GS}}\rangle=|0\rangle^{\otimes(L-4)}\otimes|\psi_{0}\rangle$
within MERA. Here, $|0\rangle\,(|1\rangle)$ is the eigenstate of
Pauli operator $\mathcal{Z}$ for each bulk spin with the higher (lower)
eigenvalue, and the states $\{|\psi_{k}\rangle\}$ for $k=0,\dots,15$
form a basis for the core region, with a Hilbert space dimension of
$2^{4}$. Specifically, $\{|\psi_{k}\rangle\}$ is chosen to be the
eigenstates of coarse-grained $H_{\partial}$ at the ``infrared''
end of MERA \cite{Vidal2007PRL,Vidal2008PRL,Evenbly2009PRB,Evenbly2016PRL,Sup_Mat},
with $k=0,\dots,15$ corresponding to increasing eigenvalues.

The hologrons, recently studied in the context of holographic models
involving MERA \cite{sahay2024arXiv,Chua2017PRB} are excited states
created by flipping bulk spins of the ground state. A generic $n$-hologron
state is defined as $|\Psi_{n}^{j_{1},\dots,j_{n}}\rangle_{h}\equiv\left(\prod_{i=1}^{n}\mathcal{X}_{j_{i}}\right)|\Psi_{\text{GS}}\rangle$,
where $\mathcal{X}_{j},\mathcal{Y}_{j},\mathcal{Z}_{j}$ are Pauli
operators acting on the $j$th bulk spin \cite{sahay2024arXiv,Chua2017PRB}.
Notably, intriguing static properties of these hologrons, such as
the existence of an attractive potential between two hologrons that
matches the prediction from AdS gravity, have been observed \cite{sahay2024arXiv}.
However, the question of their dynamical stability in the holographic
model with $H_{\partial}$ as the boundary remains unresolved. 

To address this question, we numerically study the time evolution
of the bulk system starting from a $1$-hologron state and monitor
the total probability of the bulk state $|\Psi(t)\rangle$ remaining
within the $1$-hologron subspace $\{|\Psi_{n=1}^{j}\rangle_{h}\mid j=1,\cdots,L-4\}$,
i.e., 
\begin{equation}
P_{1}^{h}(t)\equiv\sum_{j=1}^{L-4}|\langle\Psi(t)|\Psi_{n=1}^{j}\rangle_{h}|^{2}.
\end{equation}
Fig.~\ref{fig: hologron_stability}(b) shows the dynamics of $P_{1}^{h}(t)$
for two initial states, $|\Psi_{n=1}^{j=7}\rangle_{h}$ and $|\Psi_{n=1}^{j=4}\rangle_{h}$,
in a bulk system corresponding to $L=16$ sites on the boundary. We
observe that, in both cases, $P_{1}^{h}(t)$ rapidly decays from unity
during time evolution. By fitting an exponential decay function, $Ae^{-t/\tau}$,
to the numerical data in the time interval $[0,1]$, we extract lifetimes
for the single hologron: $\tau=0.51$ for $|\Psi_{n=1}^{j=7}\rangle_{h}$
and $\tau=0.85$ for $|\Psi_{n=1}^{j=4}\rangle_{h}$. These findings
confirm that hologrons are not stable excitations in the bulk system
with $H_{\partial}$ as its holographic boundary. Further detailed
checks on the stability of hologrons (see \cite{Sup_Mat}) also confirm
this conclusion. 

\begin{figure}
\includegraphics[totalheight=1.4in]{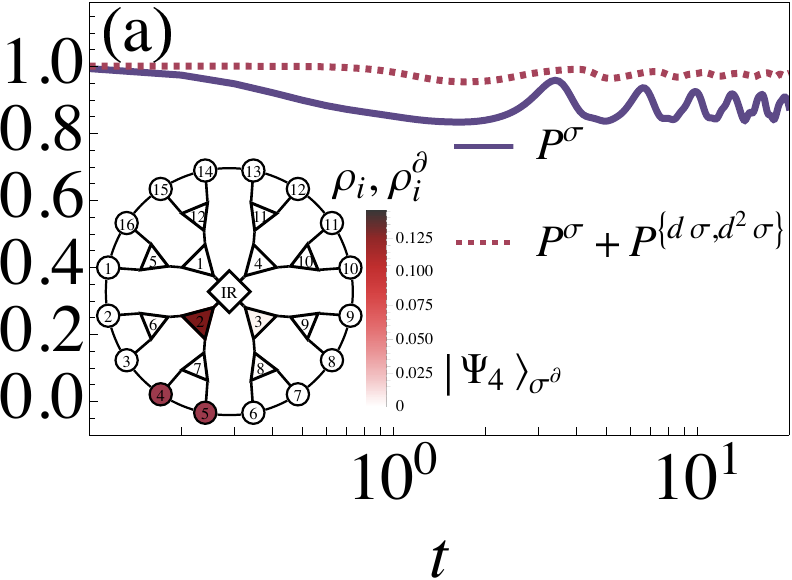}~\includegraphics[totalheight=1.4in]{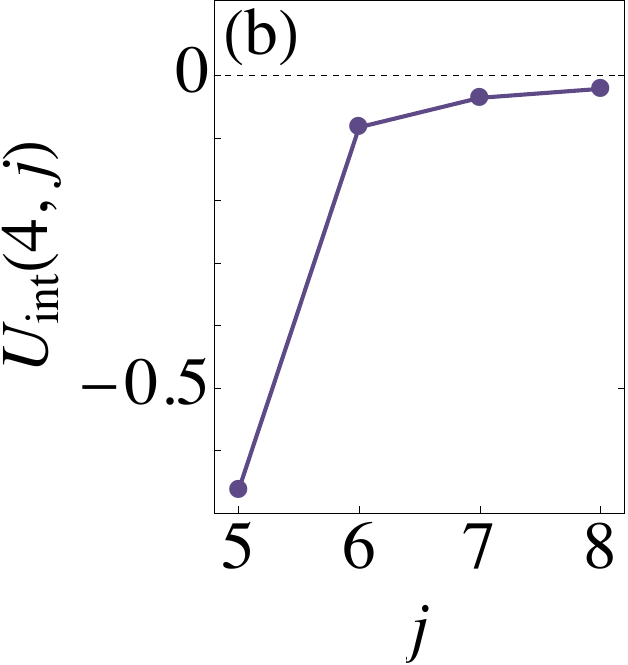}\caption{\label{fig: StableQuasiparticlesPrimary}Dynamics and interactions
of stable quasiparticles. (a) Time evolution of subspace probabilities
for the initial boundary state $|\Psi_{4}\rangle_{\sigma^{\partial}}$.
The inset shows the excitation densities $\rho_{i}$ and $\rho_{i}^{\partial}$
for $|\Psi_{4}\rangle_{\sigma^{\partial}}$. The color bar represents
the value of $\varrho_{i}$ in the bulk and $\rho_{i}^{\partial}$
on the boundary. (b) Interaction energies between quasiparticles $|\Psi_{4}\rangle_{\sigma^{\partial}}$
and $|\Psi_{j}\rangle_{\sigma^{\partial}}$ for $j=5,6,7,8$. }
\end{figure}

\emph{Stable massive quasiparticles construction via boundary CFT
primary operators.}---Interestingly, Chua et al. \cite{Chua2017PRB}
investigated a holographic model based on the MERA tensor network
with a critical Ising chain as the holographic boundary, described
by the Hamiltonian $H_{\text{C-Ising}}\equiv-L/(4\pi)\sum_{j=1}^{L}\left(Z_{j}+X_{j}X_{j+1}\right)$,
and demonstrated that the single hologron excitation is stable in
this setup. A closer examination of the hologron excitation in the
holographic model from \cite{Chua2017PRB} reveals that the single-hologron
subspace in the bulk is holographically mapped to the subspace on
the boundary generated by single spin-flip operators $X_{j}$ acting
on the ground state. These operators serve as the leading-order lattice
approximation of the primary operators in Ising CFT \cite{Zou2020PRL}.
Through the operator-state correspondence in CFT \cite{francesco2012Springer},
it follows that states generated by these spin-flip operators on the
boundary are eigenstates and, thus, stable. 

In contrast, for the holographic model considered here, the boundary
system and the MERA tensor network differ from those in \cite{Chua2017PRB}.
Consequently, the stability of the single-hologron subspace is not
guaranteed, as its instability becomes evident during dynamical evolution
{[}see Fig.~\ref{fig: hologron_stability}(b){]}. However, the above
discussion suggests that the search for stable excitations in the
bulk can be greatly facilitated by leveraging the local primary operators
of the boundary system, a construction we will carry out in the following.

Although the boundary system $H_{\partial}$ differs from $H_{\text{C-Ising}}$,
both systems are described by the same Ising CFT in the low-energy
limit \cite{Verresen2021PRX,Jones_J_Stat_Phys_2019}. This suggests
that we can construct stable boundary excitations using the primary
operators of the Ising CFT. Specifically, we focus on the spin primary
operator $\sigma^{\text{CFT}}$, which has the smallest non-trivial
scaling dimension \cite{francesco2012Springer}. Its lattice realization,
$\sigma^{\partial}$, for $H_{\partial}$ can in principle be constructed
using the variational method in \cite{Zou2020PRL}. Nevertheless,
we directly utilize the unitary map $V\equiv\prod_{j}(iX_{j}X_{j+1}-I_{j}I_{j+1})/\sqrt{2}$
that relates $H_{\partial}$ and $H_{\text{C-Ising}}$ ($H_{\partial}=V^{\dagger}H_{\text{C-Ising}}V$)
\cite{Evenbly2016PRL}, together with the lattice realization of $\sigma^{\text{CFT}}$
for $H_{\text{C-Ising}}$ \cite{Zou2020PRL}, and are able to obtain
\begin{equation}
\sigma_{j}^{\partial}=V^{\dagger}[0.635(X_{j}+X_{j+1})-0.026(X_{j}Z_{j+1}+Z_{j}X_{j+1})]V,\label{eq: sigma_bgy}
\end{equation}
where the subscript $j$ specifies the lattice site. The corresponding
single bulk excitation is then given by $|\Psi_{j}\rangle_{\sigma^{\partial}}=\mathcal{U}^{\dagger}\text{\ensuremath{\sigma_{j}^{\partial}}}\mathcal{U}|\Psi_{\text{GS}}\rangle$.
To visualize this type of quasiparticles, we compute the bulk ``excitation
densities'' $\rho_{i}\equiv\langle\Psi|(\mathcal{I}+\mathcal{Z}_{i})/2|\Psi\rangle-\langle\Psi_{\mathrm{GS}}|(\mathcal{I}+\mathcal{Z}_{i})/2|\Psi_{\mathrm{GS}}\rangle$
and its boundary counterpart $\rho_{i}^{\partial}\equiv\langle\Phi^{\partial}|(I+Z_{i})/2|\Phi^{\partial}\rangle-\langle\Phi_{\text{GS}}^{\partial}|(I+Z_{i})/2|\Phi_{\text{GS}}^{\partial}\rangle$
with $|\Phi_{\text{GS}}^{\partial}\rangle=\mathcal{U}|\Psi_{\mathrm{GS}}\rangle$
and $|\Phi^{\partial}\rangle=\mathcal{U}|\Psi\rangle$. As shown in
the inset of Fig.~\ref{fig: StableQuasiparticlesPrimary}(a), the
excitation density for $|\Psi_{4}\rangle_{\sigma^{\partial}}$ is
concentrated on a few bulk sites, indicating these quasiparticles
are localized excitations in the AdS bulk. 

From the inset of Fig.~\ref{fig: StableQuasiparticlesPrimary}(a),
it is evident that the quasiparticles constructed here receive notable
contributions from single-site flipped bulk states, or hologrons.
Thus, the mass of the hologrons, as identified in \cite{sahay2024arXiv},
is expected to confer mass to the quasiparticles constructed here.
Since massive particles in AdS geometry are known to attract one another,
it is reasonable to expect similar behavior for these quasiparticles.
To validate this, we directly calculate the interaction energy $U_{\mathrm{int}}(j,j^{\prime})$
between two bulk quasiparticles generated by two boundary spin primary
operators $\sigma_{j}^{\partial}$ and $\sigma_{j^{\prime}}^{\partial}$,
i.e., $U_{\mathrm{int}}(j,j^{\prime})=\,_{\sigma^{\partial}}\langle\Psi_{j,j^{\prime}}|H|\Psi_{j,j^{\prime}}\rangle_{\sigma^{\partial}}-\,_{\sigma^{\partial}}\langle\Psi_{j}|H|\Psi_{j}\rangle_{\sigma^{\partial}}-\,_{\sigma^{\partial}}\langle\Psi_{j^{\prime}}|H|\Psi_{j^{\prime}}\rangle_{\sigma^{\partial}}$,
with $|\Psi_{j,j^{\prime}}\rangle_{\sigma^{\partial}}\equiv\mathcal{U}^{\dagger}\text{\ensuremath{\sigma_{j}^{\partial}}}\sigma_{j^{\prime}}^{\partial}\mathcal{U}|\Psi_{\text{GS}}\rangle$
representing the two-quasiparticle state. As shown in Fig.~\ref{fig: StableQuasiparticlesPrimary}(b),
all interaction energies are negative, indicating that these quasiparticles
exhibit attractive interactions, consistent with the behavior of massive
particles in AdS geometry. 

While quasiparticles constructed from the primary operators of Ising
CFT are stable in the thermodynamic limit, finite-size effects in
experimentally realizable holographic models \cite{sahay2024arXiv}
can influence their stability. To examine this, we numerically simulate
the real-time dynamics of the bulk system initialized in a single
quasiparticle state $|\Psi_{4}\rangle_{\sigma^{\partial}}$, i.e.,
$|\Psi(t)\rangle=e^{-iHt}|\Psi_{4}\rangle_{\sigma^{\partial}}$, and
monitor the probability $P^{\sigma}(t)$, which is the total probability
of the bulk state remaining within the subspace spanned by all single-quasiparticle
states $|\Psi_{j}\rangle_{\sigma^{\partial}}$, i.e., $\mathrm{span}(\{|\Psi_{j}\rangle_{\sigma^{\partial}}|j=1,2,\cdots,L\})$.
As shown in Fig.~\ref{fig: StableQuasiparticlesPrimary}(a), the
late-time probability within this subspace is significantly enhanced
compared to the case of the hologrons {[}see Fig.~\ref{fig: hologron_stability}(b)
for comparison{]} However, a decay of approximately $12\%$ of $P^{\sigma}(t)$
is observed, which we attribute to the finite-size effects that cause
the boundary system $H^{\partial}$ to deviate from the low-energy
Ising CFT description. To further analyze these deviations, we include
the bulk subspace holographically mapped to the boundary subspace
generated by the first and second derivative descendants, ($\partial_{x}\sigma^{\text{CFT}},\partial_{\tau}\sigma^{\text{CFT}}$)
and ($\partial_{x}^{2}\sigma^{\text{CFT}},\partial_{\tau}^{2}\sigma^{\text{CFT}},\partial_{\tau}\partial_{x}\sigma^{\text{CFT}}$),
of the primary $\sigma^{\text{CFT}}$ \cite{francesco2012Springer,Zou2020PRL},
and calculate the total probability within this subspace $P^{\{d\sigma,d^{2}\sigma\}}$
(see \cite{Sup_Mat} for the explicit forms of the lattice realization
of the descendant operators and $P^{\{d\sigma,d^{2}\sigma\}}$). As
shown in Fig.~\ref{fig: StableQuasiparticlesPrimary}(a), the probability
$P^{\sigma}+P^{\{d\sigma,d^{2}\sigma\}}$ reaches approximately $97.5\%$
at late times. This suggests that couplings between subspaces generated
by spin primary operators and their descendants, absent in the Ising
CFT Hamiltonian, emerge in the finite-size boundary system $H^{\partial}$,
largely accounting for the observed decay of $|\Psi_{4}\rangle_{\sigma^{\partial}}$. 

\begin{figure}
\includegraphics[width=3in]{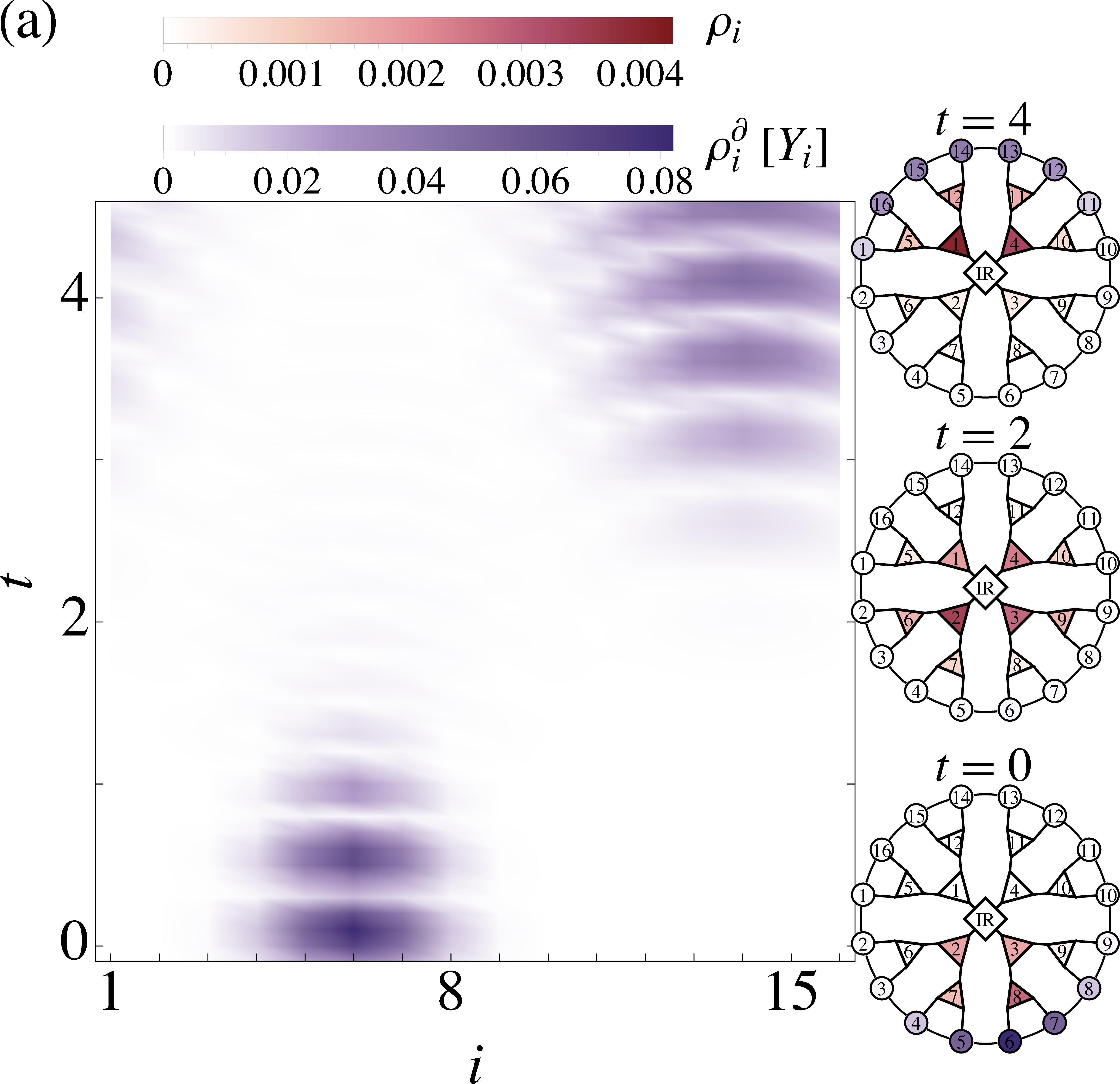}

\includegraphics[width=3.3in]{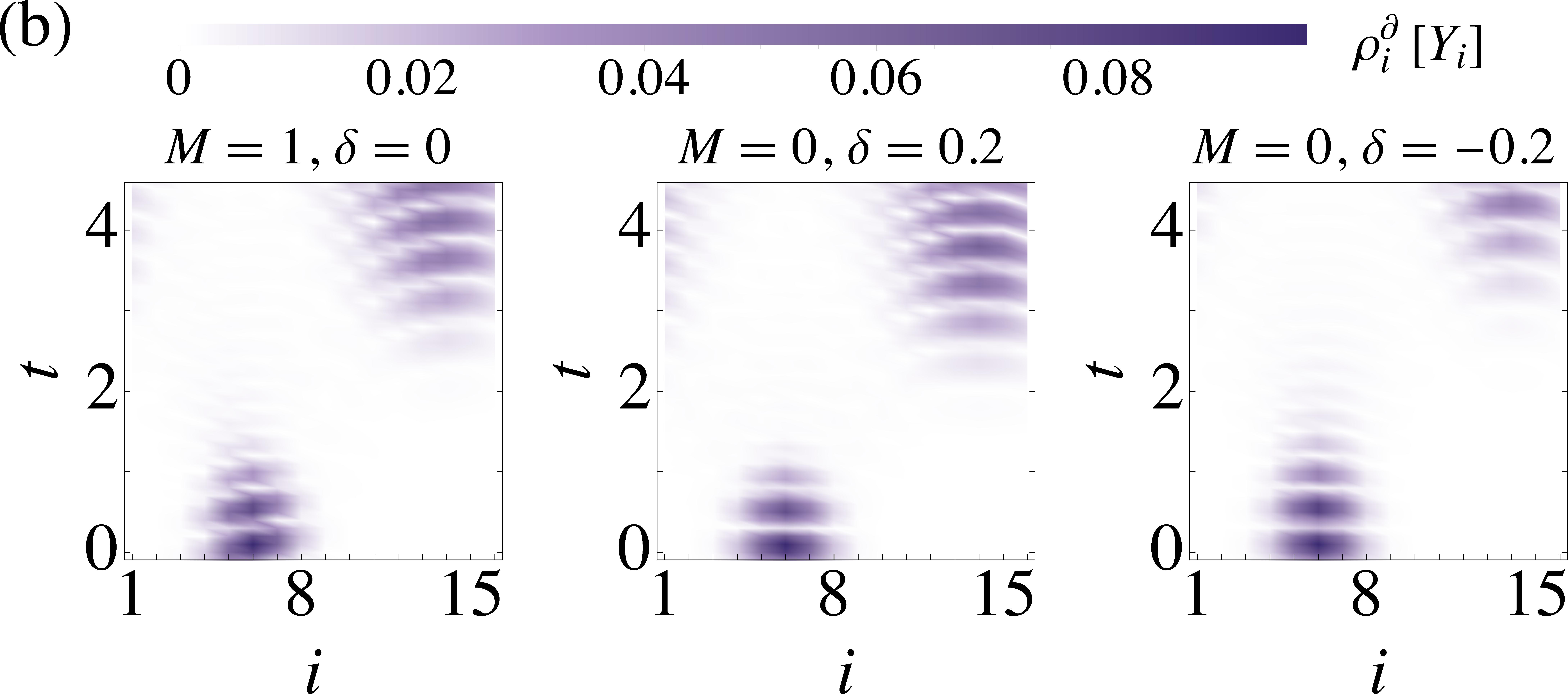}\caption{\label{fig: HoloTransport}Holographic transportation of boundary
excitations. (a) Dynamics of the ground state perturbed by Eq.~(\ref{eq: Pert_Boundary_H})
at $M=0$. The inset shows excitation densities $\rho_{i}$ and $\rho_{i}^{\partial}[Y_{i}]$
at times $t=0,2,4$. The color bar indicates the value of $\varrho_{i}$
in the bulk and $\rho_{i}^{\partial}[Y_{i}]$ on the boundary. (b)
Dynamics of the ground state perturbed by Eq.~(\ref{eq: Pert_Boundary_H})
under a Hamiltonian detuned from the critical point by an additional
$\Delta$-controlled term for $(M,\Delta)=(1,0)$, $(0,0.2)$, $(0,-0.2)$.
All other parameters in (a,~b) are set to $\Omega=5$, $\sigma_{t}=\sigma_{\phi}=0.4$.
See text for more details. }
\end{figure}

\emph{Holographic transportation of boundary excitations via AdS geodesics}.---The
above discussion demonstrated how the holographic map between the
bulk and boundary facilitates the construction of stable quasiparticles
in the bulk using the primary operators of boundary CFT. Here, we
turn our attention to another aspect of this map: how the supposed
AdS geometry of the bulk influences distinctive boundary physics. 

Among many characteristic properties of the spacetime geometry, geodesics
hold a fundamental role. Recent studies on wave packets in the context
of AdS/CFT correspondence \cite{Terashima2024PRD,Kinoshita2023JHEP}
suggest that perturbations in the boundary CFT can generate wave packets
traveling along null geodesics in the AdS bulk. Specifically, the
boundary value of the wave amplitude in the bulk is determined, to
leading order, by the sum of the perturbation strength and the corresponding
response strength (excitations on the boundary) \cite{Banks_arXiv_1998,Terashima2024PRD,Kinoshita2023JHEP,Gubser_PLB_1998,witten1998ATMP}.
This relationship implies that boundary excitations vanish as the
wave propagates through the AdS bulk but reappear when the wave reaches
the boundary again. In other words, the boundary excitations undergo
a process akin to being ``transported'' through the extra dimension
of the holographic AdS bulk, a phenomenon we term \textquotedblleft holographic
transportation.\textquotedblright{} Notably, the recently discovered
``spacetime-localized response'' \cite{bamba2024PRD}, derived via
linear response theory in the critical Ising model, corresponds to
this form of holographic transportation. 

Since the holographic map $\mathcal{U}$ between the bulk and boundary
is explicitly constructed in the holographic model studied here, we
can directly observe the \textquotedblleft hidden\textquotedblright{}
dynamics in the bulk during the holographic transportation of boundary
excitations. Fig.~\ref{fig: HoloTransport} illustrates typical holographic
transportation dynamics for boundary excitations under different system
parameters. Specifically, boundary excitations are generated by perturbing
the ground state $|\Phi_{\partial}^{\text{GS}}\rangle$ of the boundary
system using a space-time localized source $\mathcal{J}_{i}(t)$ that
couples to $Y_{i}$ on the boundary, i.e., the perturbation on the
boundary $\delta H_{\partial}(t)$ assumes the form 
\begin{equation}
\delta H_{\partial}(t)=-\sum_{i=1}^{L}\mathcal{J}_{i}(t)Y_{i},\label{eq: Pert_Boundary_H}
\end{equation}
where $\mathcal{J}_{i}(t)=A\exp\left[-(t^{2}/2\sigma_{t}^{2})-(\phi_{i}^{2}/2\sigma_{\phi}^{2})-i\Omega t+iM\phi_{i}\right]$
with $\phi_{i}=2\pi i/L-3\pi/4$. Here, $\Omega$ and $M$ correspond
to the energy and angular momentum of the geodesics in the bulk, respectively.
The perturbation amplitude $A$ is chosen as $A=\sqrt{2L/(\sigma_{t}\sigma_{\phi})}/(40\pi)$,
and $\sigma_{t}$ and $\sigma_{\phi}$ control the temporal and spatial
widths of the perturbation. To track the boundary dynamics, we define
the excitation density for $Y_{i}$ as $\rho_{i}^{\partial}[Y_{i}](t)\equiv\langle\Phi^{\partial}(t)|Y_{i}|\Phi^{\partial}(t)\rangle-\langle\Phi_{\text{GS}}^{\partial}|Y_{i}|\Phi_{\text{GS}}^{\partial}\rangle$. 

As shown in Fig.~\ref{fig: HoloTransport}(a), for the perturbation
$\delta H_{\partial}(t)$ with parameters $(\Omega,\sigma_{t},\sigma_{\phi},M)=(5,0.4,0.4,0)$,
the boundary excitation emerges at $t\sim0$, quickly vanishes at
$t\sim1.5$, and reappears at $t\sim2.5$ (see the excitation density
distribution $\rho_{i}^{\partial}[Y_{i}]$). Interestingly, during
the interval when the boundary excitation vanishes, the excitation
in the bulk propagates radially inward away from the boundary {[}middle
inset of Fig.~\ref{fig: HoloTransport}(a){]}. And when this bulk
excitation reaches the opposite side of the boundary {[}top inset
of Fig.~\ref{fig: HoloTransport}(a){]}, the boundary excitation
re-emerges. This directly demonstrates that the boundary excitation
undergoes holographic transportation through the extra dimension of
the dual AdS bulk. 

\emph{Experimental observability}.---State-of-the-art quantum simulation
platforms, such as trapped ions, superconducting qubits, and programmable
Rydberg atom arrays \cite{Moses_PRX_2023,haghshenas_PRL_2024,Arute_Nature_2019,Bernien_Nature_2017,Keesling2019nature,Ebadi2021nature,Scholl2021nature,Semeghini_Science_2021,Bluvstein_2022_nature},
have made it feasible to implement MERA tensor networks for simulating
holographic models in experimental settings. Specifically, using established
experimental protocols \cite{sahay2024arXiv} designed for Rydberg
atom arrays, boundary states corresponding to the stable bulk quasiparticles
constructed in this work---via the boundary CFT primary operators---can
be generated. The dynamics of these quasiparticles can also be observed
experimentally through time evolution under the boundary Hamiltonian,
followed by bulk-state reconstruction using a reverse MERA operation
implemented with two-qubit gates \cite{sahay2024arXiv}.

Moreover, holographic transportation can be studied experimentally
by introducing spacetime localized perturbations to the boundary system
{[}see Eq.~(\ref{eq: Pert_Boundary_H}){]}, for instance, using focused
laser pulses. The resulting dynamics can be observed by monitoring
the excitation density distribution through fluorescence signals.
Notably, holographic transportation exhibits robustness against possible
experimental imperfections. As shown in Fig.~\ref{fig: HoloTransport}(b),
this phenomenon persists even under two types of deviations: (i) perturbations
that introduce nonzero angular momentum $M$, or (ii) a boundary Hamiltonian
slightly detuned from the critical point, assuming the form $H_{\partial}-\Delta\frac{L}{4\pi}\sum_{j=1}^{L}X_{j}X_{j+1}$
with $\Delta\neq0$. In both cases, holographic transportation remains
clearly observable, demonstrating its resilience under realistic experimental
conditions. 

\emph{Conclusions}.---We have demonstrated that within the MERA model
for AdS/CFT proposed in \cite{sahay2024arXiv}, while hologrons constructed
via flipping bulk spins are dynamically unstable, stable bulk quasiparticles
with attractive interactions can be constructed using the primary
operators of the boundary Ising CFT. These quasiparticles offer a
promising avenue for investigating open questions related to gravitational
dynamics in the bulk. Additionally, we have identified a class of
boundary excitations in this MERA model that exhibit holographic transportation---a
phenomenon rooted in the AdS/CFT correspondence---where boundary
excitations are transported through the bulk's extra dimension. We
believe our findings will inspire both further theoretical exploration
of bulk dynamics in holographic models with gravitational analogs,
particularly those associated with potential black holes, as well
as experimental efforts to realize such holographic models and direct
observation of typical dynamics characteristic of AdS/CFT correspondence,
such as holographic transportation.

\begin{acknowledgments}
This work was supported by the National Key Research and Development
Program of China (Grant No.~2022YFA1405300), the National Natural
Science Foundation of China (Grant Nos.~12074180 and 12275089), Guangdong
Basic and Applied Research Foundation (Grant Nos.~2023A1515012800),
the Innovation Program for Quantum Science and Technology (Grant No.~2021ZD0301700). 
\end{acknowledgments}


%%%%%%%%%%%%%%%%%%%%%%%%%%%%%%%%%%%%%%%%%%%%%%%%%%%%%%%%%%%%%%%%%%%%%

\clearpage

%%%%%%%%%%%%%%%%%%%%%%%%%%%%%%%%%%%%%%%%%%%%%%%%%%%%%%%%%%%%%%%%%%%%%

\onecolumngrid
\vspace{\columnsep}
\begin{center}
{\large\textbf{Supplemental Material for ``Stable excitations and
holographic transportation in tensor networks of critical spin chains''}}{\large\par}
\par\end{center}

\vspace{\columnsep}
\twocolumngrid%%%%%%%%%%%%%%%%%%%%%%%%%%%%%%%%%%%%%%%%%%%%%%%%%%%%%%%%%%%%%%%%%%%%%

%\onecolumngrid
\setcounter{equation}{0}
\setcounter{figure}{0}
\setcounter{page}{1}
\def\theequation{S\arabic{equation}}
\def\thefigure{S\arabic{figure}}

%%%%%%%%%%%%%%%%%%%%%%%%%%%%%%%%%%%%%%%%%%%%%%%%%%%%%%%%%%%%%%%%%%%%%

\section{Details for Numerical implementation}

To be self-contained, we outline the minimal procedures for implementing
the MERA toy model proposed in Ref.~\cite{sahay2024arXiv}, and provide
additional details of our numerical implementation.

We first note that the Hamiltonians and ground states in both the
bulk and the boundary are represented as follows: 
\begin{equation}
\begin{array}{rcl}
\text{bulk} & \longleftrightarrow & \text{boundary}\\
\begin{array}{r}
H\\
|\Psi_{\text{GS}}\rangle
\end{array} &  & \begin{array}{l}
H_{\partial}=\mathcal{U}H\mathcal{U}^{\dagger}\\
|\Psi_{\text{GS}}^{\partial}\rangle=\mathcal{U}|\Psi_{\text{GS}}\rangle.
\end{array}
\end{array}
\end{equation}
The unitary mapping $\mathcal{U}$ is determined by
\begin{equation}
\mathcal{U}=\prod_{\ell=2}^{D-1}\left(\prod_{j=1}^{2^{\ell}}u_{2j,2j+1}\right)w^{\otimes2^{\ell}},\quad(D\equiv\log_{2}L)
\end{equation}
where the analytic expressions for $u^{\dagger}$ and $w^{\dagger}$
are given by \cite{Evenbly2016PRL}
\begin{align}
u_{j,j+1}^{\dagger}= & \frac{\sqrt{3}+2}{4}I_{j}I_{j+1}+\frac{\sqrt{3}-2}{4}Z_{j}Z_{j+1}\nonumber \\
 & +\frac{i}{4}X_{j}Y_{j+1}+\frac{i}{4}Y_{j}X_{j+1},\\
w_{j,j+1}^{\dagger}= & \frac{\sqrt{3}+\sqrt{2}}{4}I_{j}I_{j+1}+\frac{\sqrt{3}-\sqrt{2}}{4}Z_{j}Z_{j+1}\nonumber \\
 & +i\frac{1+\sqrt{2}}{4}X_{j}Y_{j+1}+i\frac{1-\sqrt{2}}{4}Y_{j}X_{j+1}.
\end{align}
Both $u$ and $w$ are represented by $4$-leg tensors in the MERA
network. 

At the operator level, given a boundary Hamiltonian $H_{\partial}$
and the mapping $\mathcal{U}$, the bulk Hamiltonian is implicitly
known. Since we can equivalently simulate the bulk dynamics via the
boundary system, we do not need an explicit formula for $H$. 

At the state level, given the boundary ground state, the bulk ground
state $|\Psi_{\text{GS}}\rangle$ can be determined. Since we are
interested in the excitations above $|\Psi_{\text{GS}}\rangle$, we
must first determine the bulk spin configuration of $|\Psi_{\text{GS}}\rangle$.
The ground state is constructed as 
\begin{equation}
|\Psi_{\text{GS}}\rangle=|0\rangle^{\otimes(L-4)}\otimes|\psi_{0}\rangle,
\end{equation}
where the $L-4$ bulk spins are polarized in the state $|0\rangle$
satisfying $\mathcal{Z}|0\rangle=|0\rangle$, and the remaining $4$
bulk spins in the infrared region are described by $|\psi_{0}\rangle$.
The infrared state $|\psi_{0}\rangle$ is unknown and must be determined
in order to minimize the ground state energy, 
\begin{equation}
E_{\text{GS}}=\langle\Psi_{\text{GS}}|\mathcal{U}^{\dagger}H_{\partial}\mathcal{U}|\Psi_{\text{GS}}\rangle.
\end{equation}
To achieve this, we need to express $E_{\text{GS}}$ in terms of $|\psi_{0}\rangle$.
Due to the renormalization process of the MERA network, $|\psi_{0}\rangle$
is the infrared state of $|\Psi_{\text{GS}}^{\partial}\rangle$ after
coarse-graining. The energy can be equivalently expressed as 
\begin{align}
E_{\text{GS}} & =\langle\psi_{0}|\mathcal{A}^{D-2}\left[H_{\partial}\right]|\psi_{0}\rangle\nonumber \\
 & =\frac{L}{4\pi}\sum_{j=1}^{L}\langle\psi_{0}|\mathcal{A}^{D-2}\left[X_{j-1}Z_{j}X_{j+1}-X_{j}X_{j+1}\right]|\psi_{0}\rangle,
\end{align}
where $\mathcal{A}$ is the so-called ascending superoperator, and
$\mathcal{A}^{D-2}$ performs coarse-graining for $D-2$ iterations.
In practice, the action of $\mathcal{A}$ on an operator is computed
via tensor contractions, as illustrated in \cite{sahay2024arXiv}.
After obtaining $\mathcal{A}^{D-2}\left[H_{\partial}\right]$, which
is a $16\times16$ matrix, we can simply diagonalize it, and $|\psi_{0}\rangle$
is taken as its ground state. The remaining $15$ excited states are
labeled by $\{|\psi_{k}\rangle\}$ with $k=1,\cdots,15$. 

In our numerical calculation, the exact ground state energy of the
boundary spin chain is $\langle\Psi_{\text{GS}}^{\partial}|H_{\partial}|\Psi_{\text{GS}}^{\partial}\rangle\approx-25.9799$,
while the ground state energy from the MERA ansatz is $\langle\Psi_{\text{GS}}|\mathcal{U}^{\dagger}H_{\partial}\mathcal{U}|\Psi_{\text{GS}}\rangle\approx-25.3972$,
with a relative error of $2.24\%$. The overlap between the exact
ground state and ansatz ground state is $\left|\langle\Psi_{\text{GS}}^{\partial}|\mathcal{U}|\Psi_{\text{GS}}\rangle\right|\approx0.9525$. 

\section{More evidence for hologron's instability}

\begin{figure}
\includegraphics[width=1.6in]{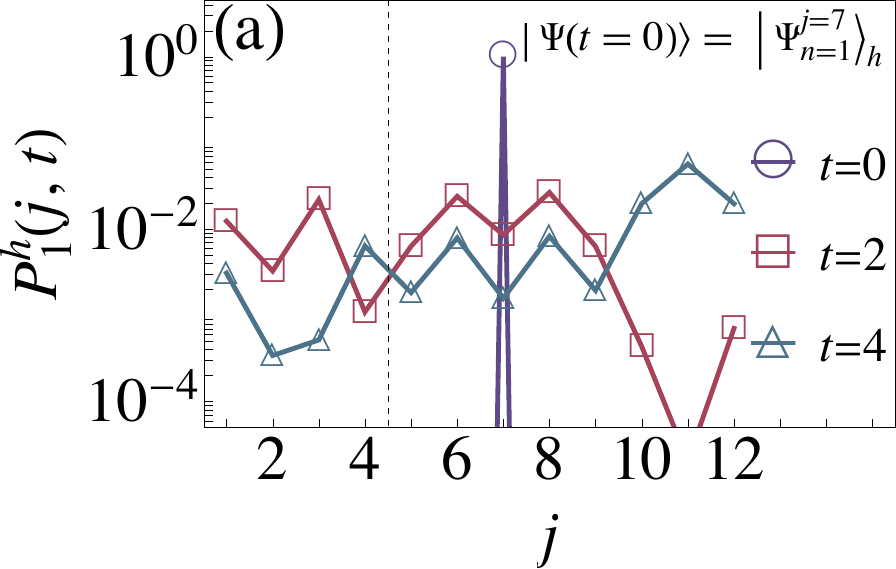}~\includegraphics[width=1.6in]{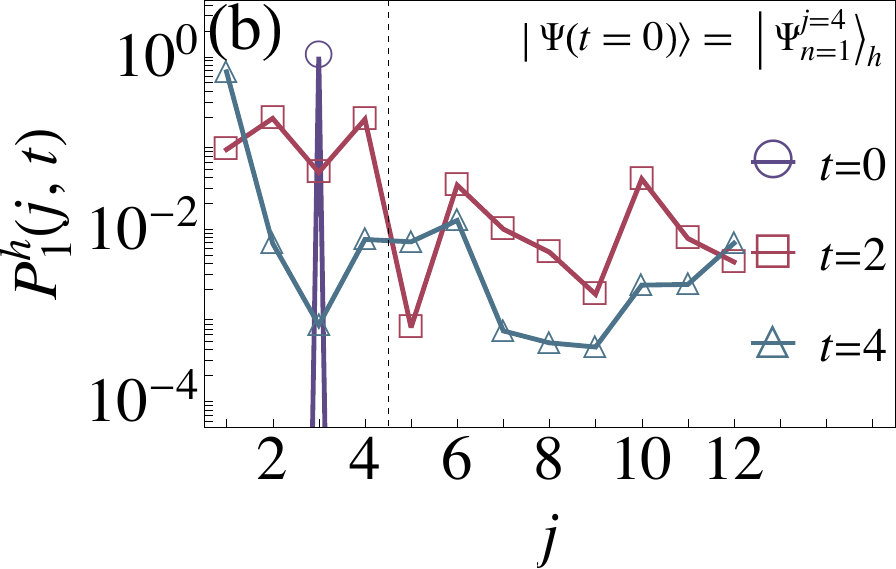}
\caption{\label{fig: dynamicsHologron}Dynamics of hologrons. (a,~b) Time
evolution of site-resolved probabilities $P_{1}^{h}(j,t)$ for initial
bulk states $|\Psi_{n=1}^{j=7}\rangle_{h}$ and $|\Psi_{n=1}^{j=4}\rangle_{h}$.
See text for more details.}
\end{figure}

Here, we provide further evidence supporting the instability of hologrons.
To investigate the behavior of a single hologron during time evolution,
we calculate the site-resolved probabilities
\begin{equation}
P_{1}^{h}(j,t)\equiv\left|\langle\Psi(t)|\Psi_{n=1}^{j}\rangle_{h}\right|^{2},
\end{equation}
where the total probability shown in Fig.~\ref{fig: hologron_stability}(b)
is given by $P_{1}^{h}(t)=\sum_{j=1}^{L-4}P_{1}^{h}(j,t)$. As illustrated
in Fig.~\ref{fig: dynamicsHologron}(a), a hologron initially located
at $j=7$ propagates across the MERA tensor network to the opposite
side, reaching sites $j=10,11,12$ with both positive and negative
angular momentum during the time interval $t=0$ to $t=4$. Similarly,
in Fig.~\ref{fig: dynamicsHologron}(b), a hologron initially at
$j=4$ moves to $j=1$, which is situated on the other side of the
network, at $t=4$. However, in both cases, the total probability
rapidly decreases by nearly an order of magnitude at $t=2$, consistent
with the findings in Fig.~\ref{fig: hologron_stability}(b). 

\begin{figure}
\includegraphics[width=1.6in]{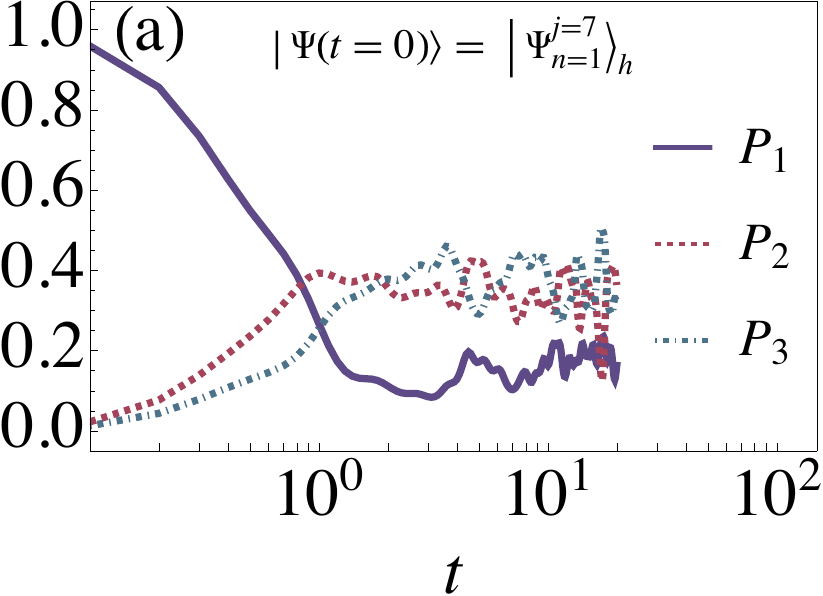}~\includegraphics[width=1.6in]{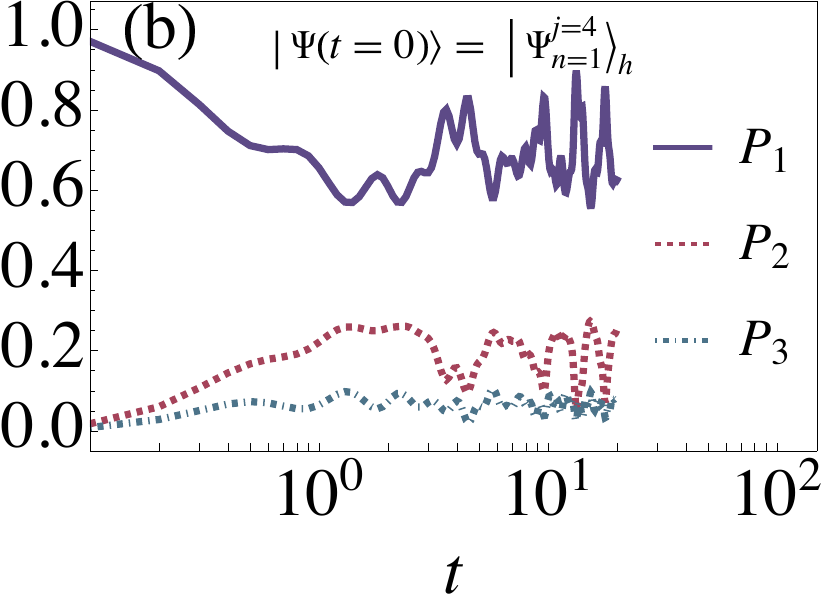}
\caption{\label{fig: generalizedstability}Instability of hologrons in the
generalized hologron subspace. (a,~b) Time evolution of subspace
probabilities for initial states $|\Psi_{n=1}^{j=7}\rangle_{h}$ (a)
and $|\Psi_{n=1}^{j=4}\rangle_{h}$ (b). }
\end{figure}

We observe that the multi-hologron states, defined by flipping bulk
spins in the ground state, cannot fully span the Hilbert space due
to the residual $2^{4}$-dimensional degrees of freedom in the infrared
region. This incomplete basis of hologron states may explain their
instability. To explore this further, we introduce a generalized hologron
state defined as 
\begin{equation}
|\Psi_{n,k}^{j_{1},\dots,j_{n}}\rangle_{h}\equiv\left(\prod_{i=1}^{n}\mathcal{X}_{j_{i}}\right)|0\rangle^{\otimes(L-4)}\otimes|\psi_{k}\rangle,
\end{equation}
where $k=0,\cdots15$ and $|\Psi_{n,k=0}^{j_{1},\dots,j_{n}}\rangle_{h}$
reduces to the original ground-state-based hologron $|\Psi_{n}^{j_{1},\dots,j_{n}}\rangle_{h}$.
Using this generalization, we simulate the dynamics and track the
total probability of the bulk state $|\Psi(t)\rangle$ remaining within
the generalized $n$-hologron subspace, defined as $\{|\Psi_{n,k}^{j}\rangle_{h}\mid j=1,\cdots,L-4;k=0,\cdots,15\}$,
i.e., 
\begin{equation}
P_{n}(t)\equiv\sum_{k=0}^{15}\sum_{j=1}^{L-4}|\langle\Psi(t)|\Psi_{n,k}^{j}\rangle_{h}|^{2}.
\end{equation}
Fig.~\ref{fig: generalizedstability} displays the dynamics of $P_{n}(t)$
for $n=1,2,3$starting from the initial bulk states $|\Psi_{n=1}^{j=7}\rangle_{h}$
and $|\Psi_{n=1}^{j=4}\rangle_{h}$. In Fig.~\ref{fig: generalizedstability}(a),
the single hologron $|\Psi_{n=1}^{j=7}\rangle_{h}$ still exhibits
rapid decay within the generalized $1$-hologron subspace, with a
lifetime $\tau=0.76$ (fitted with $Ae^{-t/\tau}$ for $0<t<1$).
Meanwhile, probabilities for higher-dimensional subspaces, such as
$P_{2}(t)$ and $P_{3}(t)$, quickly increase, indicating the decay
of a single hologron into generalized hologron pairs and triples.
Similarly, as shown in Fig.~\ref{fig: generalizedstability}(b),
the single hologron $|\Psi_{n=1}^{j=4}\rangle_{h}$ exhibits an enhanced
lifetime of $\tau=2.39$, but still decays, reaffirming its instability. 

\begin{figure}
\includegraphics[width=1.8in]{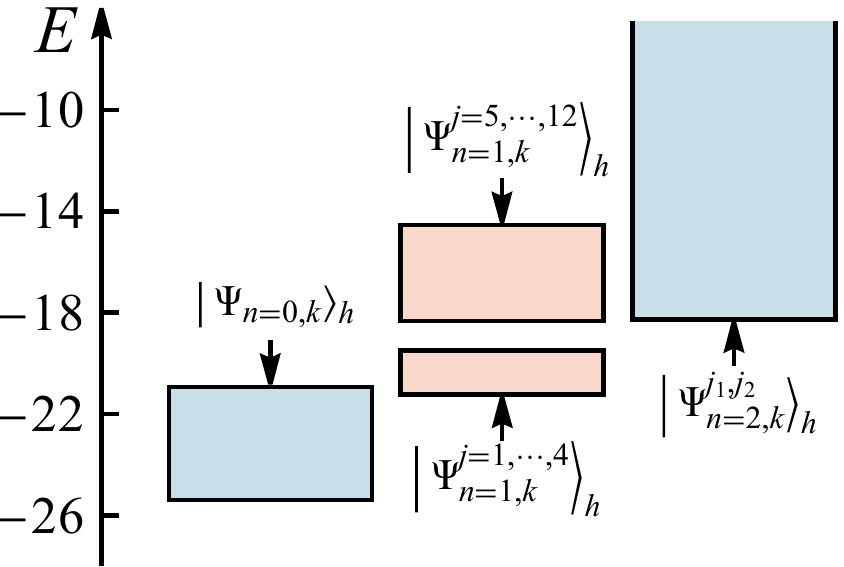}\caption{\label{fig: spectrum}Energy ranges of hologrons and generalized hologrons.
Depicted are the energy ranges for $|\Psi_{n=0,k}\rangle_{h}$, $|\Psi_{n=1,k}^{j_{1}}\rangle_{h}$,
and $|\Psi_{n=2,k}^{j_{1},j_{2}}\rangle_{h}$, with varying $j$ and
$k$. }
\end{figure}

Since unitary time evolution conserves energy, we anticipate that
the instability of hologrons will be evident in the energy spectrum.
As shown in Fig.~\ref{fig: spectrum}, we compute the energy expectation
values for all possible (generalized) $0$-, $1$- and $2$-hologron
states. Indeed, from structures of the possible energy ranges of different
generalized $n$-hologron states, one expects that the decay of a
single hologron can easily happen, since there is no obvious large
energy gap that separates the generalized $1$-hologron subspace from
other generalized multiple-hologron subspaces. 

\section{Lattice realization of derivative descendants}

From the relations provided in Table~I of \cite{Zou2020PRL}, we
observe the following approximate correspondences
\begin{align}
Y_{j}\sim & 0.8031i\partial_{\tau}\sigma^{\text{CFT}},\nonumber \\
X_{j}\sim & 0.803121\sigma^{\text{CFT}}-0.017\partial_{\tau}^{2}\sigma^{\text{CFT}}\\
 & -0.033\partial_{x}^{2}\sigma^{\text{CFT}},\\
X_{j}Z_{j+1}+Z_{j}X_{j+1}\sim & 0.803121\sigma^{\text{CFT}}-0.820\partial_{\tau}^{2}\sigma^{\text{CFT}}\nonumber \\
 & -0.736\partial_{x}^{2}\sigma^{\text{CFT}},\\
X_{j}Z_{j+1}-Z_{j}X_{j+1}\sim & 1.205\partial_{x}\sigma^{\text{CFT}},\\
Y_{j}Z_{j+1}+Z_{j}Y_{j+1}\sim & 2.41i\partial_{\tau}\sigma^{\text{CFT}},\\
Y_{j}Z_{j+1}-Z_{j}Y_{j+1}\sim & -0.4015i\partial_{\tau}\partial_{x}\sigma^{\text{CFT}}.
\end{align}
These relations suggest that the subspace consisting of the spin primary
states and their first and second derivative descendants is defined
as 
\begin{equation}
\mathcal{V}\equiv\mathrm{span}(\{\mathcal{U}^{\dagger}A_{j}\mathcal{U}|\Psi_{\text{GS}}\rangle|A_{j},j\}),
\end{equation}
where $A_{j}$ runs over the set $\{X_{j},Y_{j},X_{j}Z_{j+1},Z_{j}X_{j+1},Y_{j}Z_{j+1},Z_{j}Y_{j+1}\}$
and $j$ runs over $\{1,2,\cdots L\}$. Thus, we define the subspace
of first- and second-derivative descendants as 
\begin{equation}
\mathcal{V}/\mathrm{span}(\{|\Psi_{j}\rangle_{\sigma^{\partial}}|j\}).
\end{equation}
The probability of a state within this subspace is $P^{\{d\sigma,d^{2}\sigma\}}$
in Fig.~\ref{fig: StableQuasiparticlesPrimary}(a). 
\end{document}